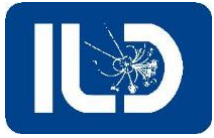

# Probing the CP properties of the Higgs sector at ILC

T. Agatonovic Jovin[1)*], I. Bozovic Jelisavcic[*], I. Smiljanic[*], G. Kacarevic[*], N. Vukasinovic[*], G. Milutinovic Dumbelovic[*], J. Stevanovic[‡], M. Radulovic[‡], D. Jeans[†]

On behalf of the ILD Concept Group

[*] *Vinca Institute of Nuclear Sciences - National Institute of the Republic of Serbia, University of Belgrade, Serbia,* [‡] *Department of Physics, Faculty of Science, Kragujevac University, Kragujevac, Serbia,* , [†] *Institute of Particle and Nuclear Studies, High Energy Accelerator Research Organization (KEK), Tsukuba, Japan*

## Abstract

The violation of CP symmetry is one of Sakharov's conditions for the matter-antimatter asymmetry of the Universe. Currently known sources of CP violation in the quark and neutrino sectors are insufficient to account for this. Is CP also violated in the Higgs sector? Could the SM-like Higgs boson be a mixture of even and odd CP states of an extended Higgs sector? With what precision could such effects be measured at future electron-positron colliders? These questions will be discussed in the light of the latest and ongoing studies at ILC.

[1)] tatjana.jovin@vinca.rs

# 1 Introduction

CP violation is a well-known problem related to numerous open questions in contemporary physics. The predicted size of CP violation (CPV) in the SM is insufficient to explain the baryon asymmetry of the Universe (BAU). $B$ and $K$-meson observed CPV is also insufficient to explain the Universe as we know it. On the other hand, even though the discovery of the Higgs boson at the LHC [1] completes the SM theory with respect to the mass generation mechanisms, profound questions related to the Higgs boson as the only fundamental scalar discovered remain: the hierarchy problem and the stabilization of the scalar masses, including the Higgs boson, the expectation value of the Higgs vacuum and the energy density of the Universe, Higgs sector as a portal to and dark matter, Higgs boson as the inflaton, and many more. We can logically ask – are these sets of open questions connected? CP violation in the Higgs sector may provide the answer.

In the SM the Higgs boson is a CP even scalar, while many extensions of the SM introduce additional Higgs bosons, often including CP odd pseudoscalar state. Higgs boson mass eigenstates could be a mixture of even and odd CP states. In many models of physics beyond the SM, a CP odd component does not couple directly to $W$ and $Z$ bosons, while the couplings to leptons are typically not suppressed and therefore provide better sensitivity to probe CP properties of the Higgs sector. Being a decay to the heaviest lepton, $H \to \tau\tau$ decays offer such an option. Yet, the Higgs-vector boson ($HVV$) vertex can be probed both in the Higgs production ($VV$-fusion) and decay ($H \to VV$) channels. Both studies are either completed ($H \to \tau\tau$) [2] or ongoing (inclusive Higgs production in ZZ-fusion). Plans and perspectives of the CPV studies at ILC are also reviewed in [3].

Electron-positron colliders, and ILC in particular as a Higgs factory, offer numerous Higgs production processes (shown on Figure 1) in an almost QCD background free experimental environment. At linear machines like ILC, capable of reaching TeV center-of-mass energies, Higgs statistics is enriched in particular by Higgs production in $WW$-fusion, while electron and positron beam polarization enhance statistical sensitivity and plays a role in background suppression.

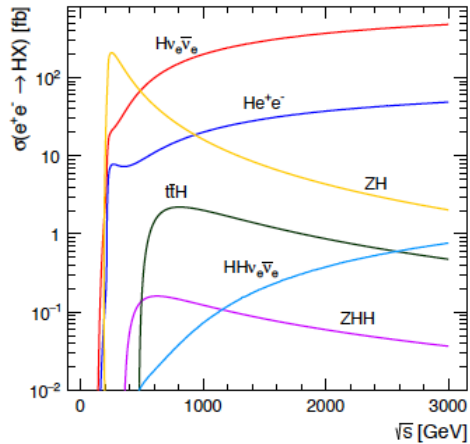

Figure 1. Higgs boson production processes at electron-positron machines [4].

# 2 Ways to probe CP violation in the Higgs sector

To study the Higgs CP properties by probing $HVV$ or $Hff$ vertices, correlation between spin orientations of the Higgs boson's decay products or parent bosons can be exploited. As illustrated in Figure 2, both Higgs production and decays can be used to kinematically recover the information of the Higgs boson CP state, offering possibilities for sensitivity enhancement of the CPV mixing angle measurement.

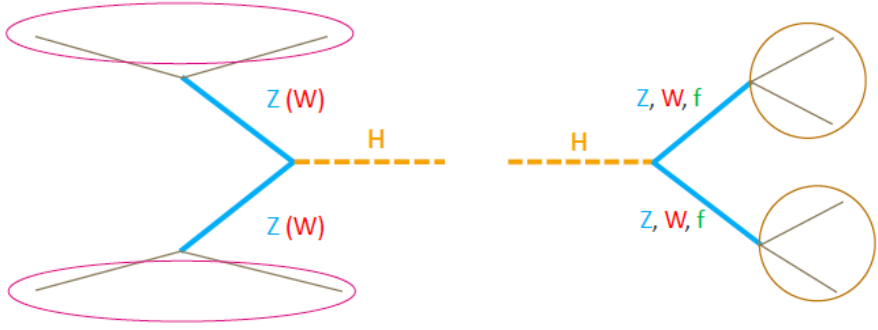

Figure 2. Both Higgs production (left) and decay vertices (right) can be exploited to probe CPV in the Higgs sector.

In $VV$-fusion or Higgs decay to $VV$, information on spin orientation of $VV$ states is contained in the angle between the production planes ($\phi$), which becomes the CPV sensitive observable. In case of ZZ-fusion process illustrated in Figure 3a, $\phi$ angle can be determined as:

$$\phi = a \arccos(\hat{n}_1 \cdot \hat{n}_2) \quad (1)$$

where the unit vectors orthogonal to the production planes are indexed with $\hat{n}_1, \hat{n}_2$ and defined as:

$$\hat{n}_1 = \frac{q_{ei^-} \times q_{ef^-}}{|q_{ei^-} \times q_{ef^-}|} \quad \text{and} \quad \hat{n}_2 = \frac{q_{ei^+} \times q_{ef^+}}{|q_{ei^+} \times q_{ef^+}|} \quad (2)$$

Coefficient $a$ is defined as a normalized triple product:

$$\frac{q_V \cdot (\hat{n}_1 \times \hat{n}_2)}{|q_V \cdot (\hat{n}_1 \times \hat{n}_2)|} \quad (3)$$

where $V$ stands for a Z boson in the electron plane or an on-shell boson in case of a Higgs decay. The value of $a$ determines how the second (positron) plane is rotated with respect to the first (electron) plane (*forward* or *backward*), while $q_{i(f)}$ stands for initial (final) state particle momentum. If the second plane falls backwards (as illustrated) $a = -1$, otherwise $a = 1$. As there are three (momentum) vectors in one production plane (initial and final state electron or positron and a Z boson), there is more than one convention to define the orthogonal unit vectors to the production planes. In Figure 3a and with convention described above, vectors $\hat{n}_1$ and $\hat{n}_2$ orthogonal to the production planes have the same direction, thus in Eq. (1), $a$ takes the positive value. Similarly, Figure 3b illustrates Higgs decay to $WW$ bosons. According to the applied convention, the unit vectors orthogonal to decay planes are now opposite. Eq. (1) takes the form:

$$\phi = a \arccos(-\hat{n}_1 \cdot \hat{n}_2) \quad (4)$$

In case of a Higgs decay, off-shell boson $V^*$ decay plane is by definition the second plane and if it falls backwards (as illustrated) $a = -1$, otherwise $a = 1$, where direction of motion of the on-shell boson ($V$) regulates the notion of direction (fwd. or back.). In case of Higgs (vector boson) decays, it is essential to distinguish between final state fermion and antifermion, using the information about the jet or lepton charge.

Figure 4 illustrates distributions of the angle $\phi$, at the generator level (WHIZARD v1.95 [5]) for Higgs decays to ZZ bosons (top left) and WW bosons (top middle) and in Higgs production via ZZ-fusion (top right) at 1 TeV ILC. The bottom row shows in red theoretical expectations [6] for the SM predictions of the corresponding $\phi$ distributions above.

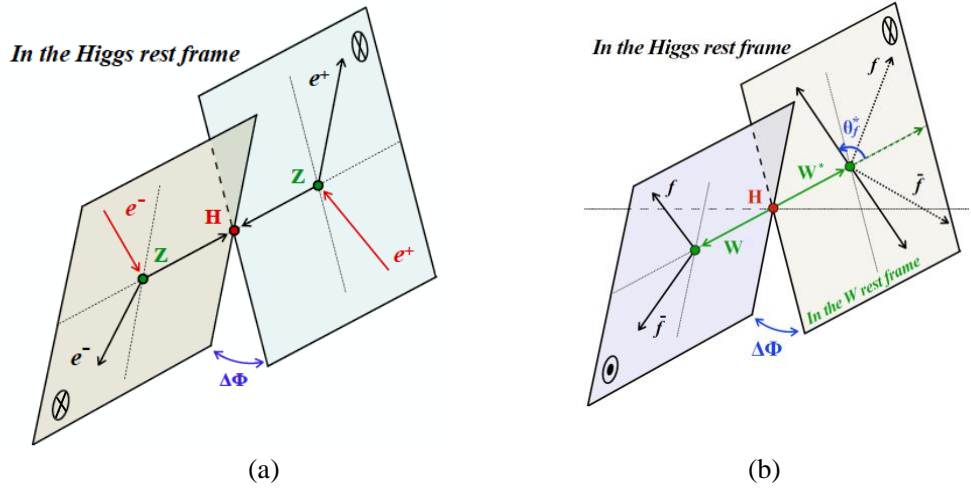

Figure 3 Schematic view of the Higgs boson production and decay planes illustrating definition of the $\phi$ angle in ZZ-fusion (a) and $H \rightarrow WW$ decay (b).

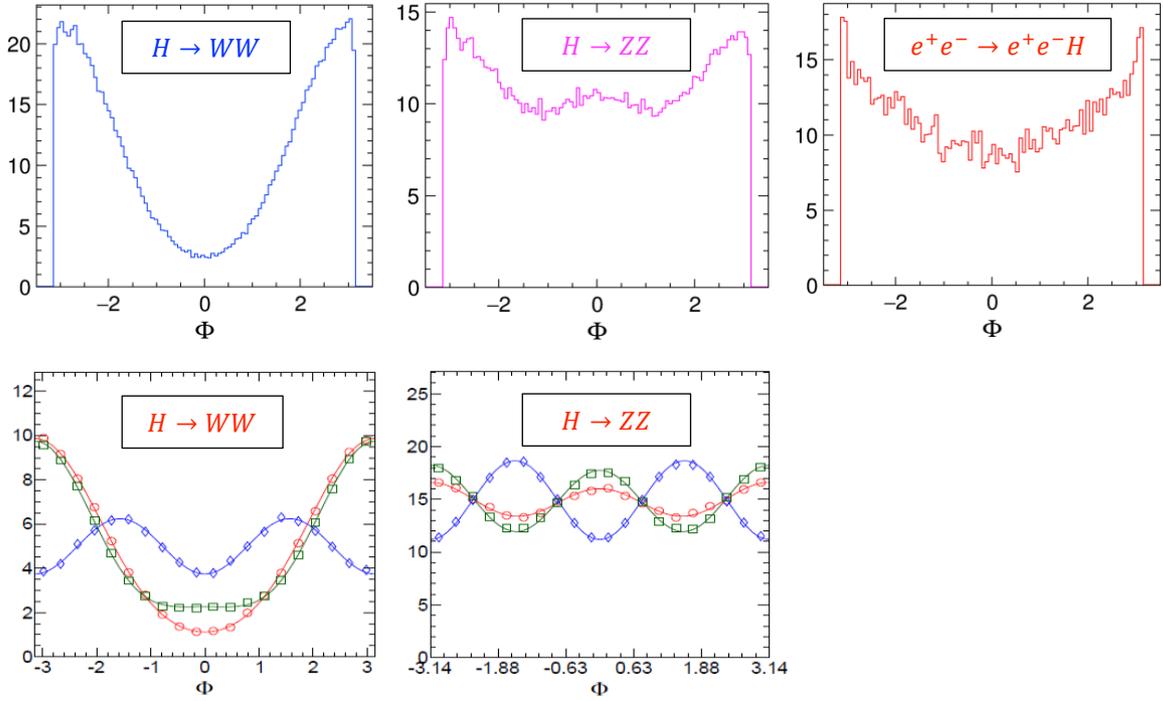

Figure 4. Distributions of the angle $\phi$ reconstructed for the simulated data (upper row) and the corresponding theoretical expectations in red (bottom row).

As already mentioned, $Hff$ vertices offer better sensitivity than $HVV$ to CPV contributions in the Lagrangian (Eq. (5)):

$$\mathcal{L}_{ffH} \sim g\bar{f}(\cos\psi_{CP} + i\gamma^5 \sin\psi_{CP})fH \qquad (5)$$

where $\psi_{CP}$ stands for the mixing angle between CP even and CP odd Higgs states. Exploiting the strongest Higgs to lepton coupling ($H\tau\tau$), uncertainty of 75 mrad (4.3 deg) is found with 2 ab$^{-1}$ of polarized data at 250 GeV [2]. This is currently the most precise projection of future projects' sensitivity to measure Higgs CPV mixing angle (i.e projected absolute uncertainties at HL-LHC and FCCee are for a factor of two (HL-LHC or more (FCCee) larger [7]. Distributions of angle $\phi$ for different values of the Higgs mixing angle $\psi_{CP}$ is given in Figure 5 for $H \rightarrow \tau\tau$ decays.

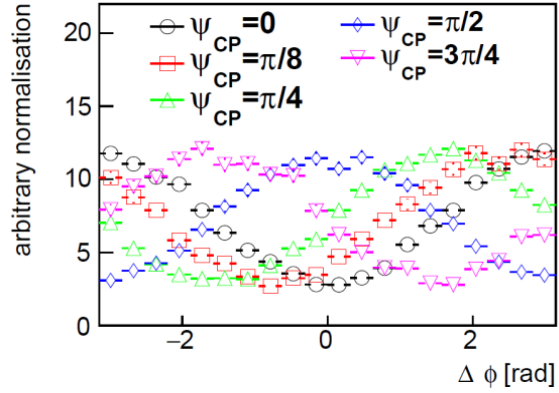

Figure 5. Distributions of angle $\phi$ reconstructed from the simulated data, in Higgs to $\tau\tau$ decays, for different values of the Higgs CPV mixing angle $\psi_{CP}$ [2].

## 3 Summary

At ILC, similarly to other electron-positron colliders, both Higgs production and decay vertices ($HVV, Hff$) can be exploited to probe the Higgs CP properties. Due to $Hff$ vertex sensitivity to CPV contribution at a tree level, Higgs decay to tau leptons is studied at 250 GeV ILC giving the utmost precision among available future projects' projections. Due to a rising cross-section for VV-fusion with a cenetr-of-mass energy both Higgs production and decays to vector bosons can be exploited. There is ongoing CPV study of the inclusively produced Higgs bosons in ZZ-fusion at 1 TeV ILC.